\documentclass[runningheads]{llncs}

\usepackage{array}
\usepackage{multirow}
\usepackage{graphicx}
\usepackage{pdfpages}
\usepackage{colortbl}
\usepackage{xcolor}
\usepackage{algorithm}
\usepackage{algpseudocode}
\usepackage{xspace}
\usepackage{amsmath}
\usepackage{subfigure}
\usepackage{amssymb}
\usepackage{enumerate}
\usepackage{tabularx} 
\usepackage{longtable} 
\usepackage{tikz}
\usetikzlibrary{shapes,arrows}
\usepackage[utf8]{inputenc}

\begin{document}

\title{
A Hybrid Approach to Monitor Context Parameters for Optimising Caching for Context-Aware IoT Applications
\thanks{Supported by Australian Research Council (ARC) Discovery Project Grant DP200102299.}}

\titlerunning{Hybrid IoT Context Caching Optimization}

%
\author{Ashish Manchanda\inst{1}\and
Prem Prakash Jayaraman\inst{1}\and
Abhik Banerjee\inst{1}\and
Arkady Zaslavsky\inst{2} \and
Shakthi Weerasinghe\inst{2} \and
Guang-Li Huang \inst{2} 
}
\authorrunning{A. Manchanda et al.}
%
\institute{Swinburne University of Technology, Melbourne, Australia \and
Deakin University, Melbourne, Australia}
\maketitle              
\begin{abstract}

Internet of Things (IoT) has seen a prolific rise in recent times and provides the ability to solve several key challenges faced by our societies and environment. Data produced by IoT provides a significant opportunity to infer context that is key for IoT applications to make decisions/actuations. Context Management Platform (CMP) is a middleware to facilitate the exchange and management of such context information among IoT applications. In this paper, we propose a novel approach to monitoring context freshness as a key metric, to improving the CMP's caching performance to support the real-time context needs of IoT applications. Our proposed hybrid algorithm uses Analytic Hierarchy Process (AHP) and Sliding Window technique to ensure the most relevant (as needed by the IoT applications) context information is cached. By continuously monitoring and prioritizing context attributes, the strategy adapts to IoT environment changes, keeping cached context fresh and reliable. Through experimental evaluation and using mock data obtained from a real-world mobile IoT scenario in section~\ref{use case}, we demonstrate that the proposed algorithm can substantially enhance context cache performance, by monitoring the context attributes in real time.

\end{abstract}
\section{Introduction}

IoT applications play a vital role in collecting and processing data from various IoT sensors to provide insights into supporting decision making~\cite{s23031711}. The data from IoT sensors is key to infer context that is required to provide relevant services to users. Context as defined by Anind Dey~\cite{dey2001understanding} is ``Any information that can be used to characterize the situation of an entity. An entity is a person, place, or object that is considered relevant to the interaction between a user and an application, including the user and applications themselves."

Context Management Platform (CMP) has been proposed in the literature
as a middleware to facilitate the exchange of context between context providers and IoT applications~\cite{hassani2018context}.

The increasing number of IoT applications and the need to provide context to meet the real-time demands of such application pose several performance challenges for CMPs. Delays in serving context to IoT applications can have severe consequences due to delays or failure in decision-making/actuation. Hence, it is a key challenge to manage the need and serve context to meet the demands.

To mitigate this delay and enhance the efficiency of the system, caching context becomes an essential strategy. By storing context information in the cache minimizes the time-consuming process of context retrieval from context providers, thus allowing for real-time access to context for IoT applications. This adaptation not only optimizes the performance of the CMP but also meets the growing demand for real-time, responsive IoT applications~\cite{perera2014context}.

Caching context for IoT applications to utilize, presents a unique challenge, distinctly different from traditional data caching. The challenges associated with caching context for IoT applications stem from the complex and dynamic characteristics of context. Traditional data, which is typically static, can be cached using well-established metrics like frequency and size~\cite{Al-Ward2022Caching}. In contrast, context is very dynamic and is continuously changing which demands more nuanced caching strategies.

A multitude of factors come into play when determining which context needs to be cached and when to refresh or discard it. As context-aware IoT applications rely on context for their operations, it's imperative to cache it efficiently to ensure quick and accurate retrieval of context. There is a need to support continuous monitoring of key metrics and a critical parameter called ``context freshness" which we introduce in this paper. This metric ensures the current state of the context inside the CMP whether the context needs to be cached anew, evicted, or left unchanged in the cache memory. In summary, due to the dynamic nature of context, and the importance of keeping it fresh require specialized strategies designed specifically for IoT applications~\cite{al2015internet,ryan2000from}. This guarantees that real-time sensitive IoT applications can make well-informed decisions, even in rapidly changing situations.

Let's consider a scenario~\label{use case} of a smart IoT navigation system in Autonomous Vehicles(AVs) in a busy city. Here, an autonomous vehicle moves smoothly using a new IoT system for navigation. This system doesn't only use the vehicle's own sensors; it also uses information from other vehicles. For example, if a car detects a "Road Work" sign or an obstacle, it sends this information to a context management platform. This information is then available for all connected vehicles. So, another car coming to the same area knows about the obstacle before reaching it, making the drive safer and more efficient. By sharing information in this way, vehicles can better handle changing road situations.

Now current autonomous vehicles~\cite{KRASNIQI2016269} often face difficulties in detecting situations such as road works, accidents, or heavy traffic congestion, because their internal models need vast amounts of training data to predict and navigate these situations effectively~\cite{Iyer2021AI}. Contextual awareness of road conditions as discussed in the scenario above can greatly help in navigation and in some cases turning off auto-pilot.
Given the ever-changing nature of such scenarios (road conditions) - with work starting or ending, lanes getting blocked or opened up - the underlying context representing the situations ``Road Work" is constant changing (dynamic).

Research in monitoring the parameters of context caching, particularly for context freshness, is in its early stages~\cite{Al-Ward2022Caching,weerasinghe2023traditional}. 
The monitoring of query logs~\cite{khargharia2022probabilistic}, as a solution has been considered in the past but this approach lacks the dynamism required to keep up with the ever-changing freshness of context that characterizes a situation like road works. This motivating scenario highlights the unique challenges posed by context caching in an IoT ecosystem in CMP, laying the groundwork for our research. 
In this paper, we propose the Context Freshness Monitoring System (CFMS) to monitor context-parameter/key-metric ``context freshness" with the aim of enhancing cache optimization for context-aware IoT applications. Our work also includes comparisons with conventional algorithms such as RU and FIFO in section~\ref{sec:evaluation}. CFMS uses two algorithms for its function that are:
\begin{enumerate}
    \item \textbf{{Decision Supporting Algorithm(DSA)}} : The main aim of this algorithm is to determine and prioritize the weights of context attributes using the Analytical Hierarchical Process (AHP) based on their relative significance. By doing so, it ensures that context-aware applications can efficiently process and act upon the most relevant context attributes in changing situations such as road work.
    \item \textbf{{Parameters Freshness Processing Algorithm(PFPA)}} : The main aim of this algorithm is to take input of prioritize context attributes from decision supporting algorithm and cache them based on their respective weights. This algorithm maintains the freshness of cached context and makes real-time decisions on updating context attributes when they exceed predefined thresholds using sliding window mechanism. By doing so, it ensures that the most relevant and recent parameters are readily available for context-aware IoT applications, thereby optimizing performance and accuracy.
\end{enumerate} 

The CFMS integrates the strengths of its sub-algorithms to determine optimal caching decisions for context-aware IoT applications. The DSA employs the Analytical Hierarchical Process (AHP) to assign weights to context attributes, thereby identifying their relative importance. Subsequently, the PFPA leverages these weights to prioritize which attributes are cached. Using a sliding window mechanism, this algorithm ensures that the cache is consistently refreshed, maintaining only the most relevant and current context attributes. Collectively, the CFMS monitors the key metric ``context freshness" which ensures that the caching mechanism is both efficient and contextually relevant. 

The remainder of the paper is as follows. Section \ref{sec:literature} reviews the Related Work. Section~\ref{sec:Design} describes the design and implementation of the Context Freshness Monitoring System (CFMS). Section~\ref{sec:evaluation} evaluates the CFMS model. Section~\ref{sec:discussion} discusses and concludes the paper.

\section{Related Work}~\label{sec:literature}

Research in the field of IoT has recognized the importance of data caching for improving system performance. Traditionally, data caching strategies have relied on parameters like popularity and logs to enhance cache performance~\cite{weerasinghe2023traditional}. However, due to the static nature of such data and the rapidly changing IoT environment, these approaches may not always provide the most relevant or updated information~\cite{rizwan2016realtime,bettini2010context}. Whereas context in IoT refers to the situational, environmental, or operational information that can affect the behavior or interpretation of IoT devices or data. For example, in a context-aware smart home, the context could include whether the resident is home or away, time of the day, day of the week, current weather conditions, etc. Context data involves a lot of factors and attributes that can significantly affect the operation of IoT devices and applications.

In recent advancements concerning context caching within IoT applications, a significant contribution was made in the paper~\cite{khargharia2022probabilistic}. This work focuses on the optimization of a context management platform (CMP), named CoaaS. Their novel system, ConCaf, utilizes context query logs and machine learning techniques to estimate the demand probability of context information. This approach is fundamentally different from our method, as their primary metric for caching decisions is derived from monitoring query logs and optimizing the context caching probability by adjusting the hyperparameters of the regression models. Although the results of their evaluation showed a significant improvement in the response time of CoaaS, the emphasis was on the proactive placement of context information based on query demand. In contrast, our work specifically zeroes in on the real-time monitoring of context attributes, aiming to maintain a context freshness metric inside the CMP. This distinction underscores our approach's commitment to addressing the transient and dynamic nature of contexts in IoT environments, ensuring that cached data remains relevant and timely.

Context-aware systems consider a multitude of context attributes that provide a high-level interpretation of data and allow systems to make intelligent decisions~\cite{fernandez2019realtime,shekade2020vehicle}. Nevertheless, the transient nature of these parameters introduces challenges in maintaining updated cache which, in turn, could create potential bottlenecks in system performance.

Several studies have explored adaptive caching strategies for context-aware systems~\cite{10.1145/3558053.3558057}, but they often assume that all cached context data is equally relevant. Such approaches do not take into account the freshness of the context or the real-time changes in context parameters. Despite numerous research studies focusing on the need for context-awareness in IoT systems, comprehensive studies emphasizing real-time monitoring of context parameters to maintain context freshness are limited~\cite{stankovic2014research,gu2005middleware}. 

There has been a good amount of literature, that predominantly investigates various techniques and technologies employed in real-time traffic monitoring across diverse contexts. Utilizing ontologies, frameworks like CARE, Wireless Sensor Networks, the fusion of visual perceptions with Convolutional Neural Networks (CNN), and IoT, these studies explore traffic evaluation, collision detection, and traffic parameter monitoring~\cite{ref_article1,ref_article2,ref_article3,ref_article4,ref_article5,ref_article6,ref_article7}. However, a conspicuous gap lies in these studies' attention to the crucial aspects of context caching and context freshness within IoT applications. Despite the substantial progress made in real-time decision-making for IoT applications, a comprehensive approach towards maintaining the context freshness and effective context caching strategies in IoT systems remains largely unexplored.

The algorithm discussed in the study efficiently incorporates context-specific content popularity to improve cache hits~\cite{Muller2017}. However, its reliance on content popularity as the main parameter can limit its effectiveness in IoT applications, where a plethora of context attributes and factors define the context. Our research seeks to address this gap by focusing on real-time monitoring of multiple context parameters.

The studies explore content caching strategies in IoT, focusing on balancing energy consumption and content freshness, and optimizing UAV-based network performance considering spatial parameters~\cite{XU201924,app8101959}. However, they overlook the potential of continuous real-time monitoring of diverse context parameters. Our approach enhances these strategies by incorporating comprehensive context parameter monitoring, leading to nuanced insights into context caching. This could significantly improve decision-making, system performance, and reliability in IoT applications.

Thus, there is a clear need for research that not only considers context caching but also addresses the dynamic nature of context parameters. This gap in literature justifies the necessity for our research, focusing on monitoring these context attributes/parameters and the implications on metric ``context freshness" as a critical parameter in real-time IoT applications.

\section{Context Freshness Monitoring System(CFMS)}~\label{sec:Design}

In the realm of Internet of Things (IoT), the term ``context" refers to any information that can be used to understand the situation of an entity. An entity can be a person, place, or object considered relevant to the interaction between a user and an application, including the user and the application themselves~\cite{dey2001understanding}. Each building block of this information is termed as a ``context attribute".

Taking the IoT ecosystem with autonomous vehicles as an illustrative example, let's delineate the concepts:

\begin{enumerate}
    \item \textbf{Context Attribute}: These are individual pieces of information, or parameters, that can provide insights about a specific aspect of the environment(situation)~\cite{padovitz2004theory,boytsov2011sensory}. For contextual information such as ``road work" as discussed in the motivating scenario, context attributes could be ``speed of vehicle", ``weather conditions", ``traffic density", or ``road quality". Each attribute offers insight into a specific aspect of the driving conditions.

    \item \textbf{Situation/Context}: This represents a broader, composite understanding that is derived from the combination of various context attributes. For autonomous vehicles, a ``road work" situation would be an amalgamation of multiple context attributes: a decrease in vehicle speed, the presence of roadblocks or diversions, detection of construction machinery, and maybe an increase in dust levels ~\cite{padovitz2004theory,boytsov2011sensory}.
\end{enumerate}

The dynamism of these context attributes, especially in fluctuating scenarios such as road conditions, underscores the importance of ``context freshness". Ensuring that the context attributes being considered are the most recent and relevant is crucial for accurate contextual awareness, particularly in real-time environments like autonomous driving. Hence, it becomes paramount to continually monitor and update the context cache, addressing the unique challenges of context freshness in IoT ecosystems.

\begin{figure}
\centering
\includegraphics[width=12cm]{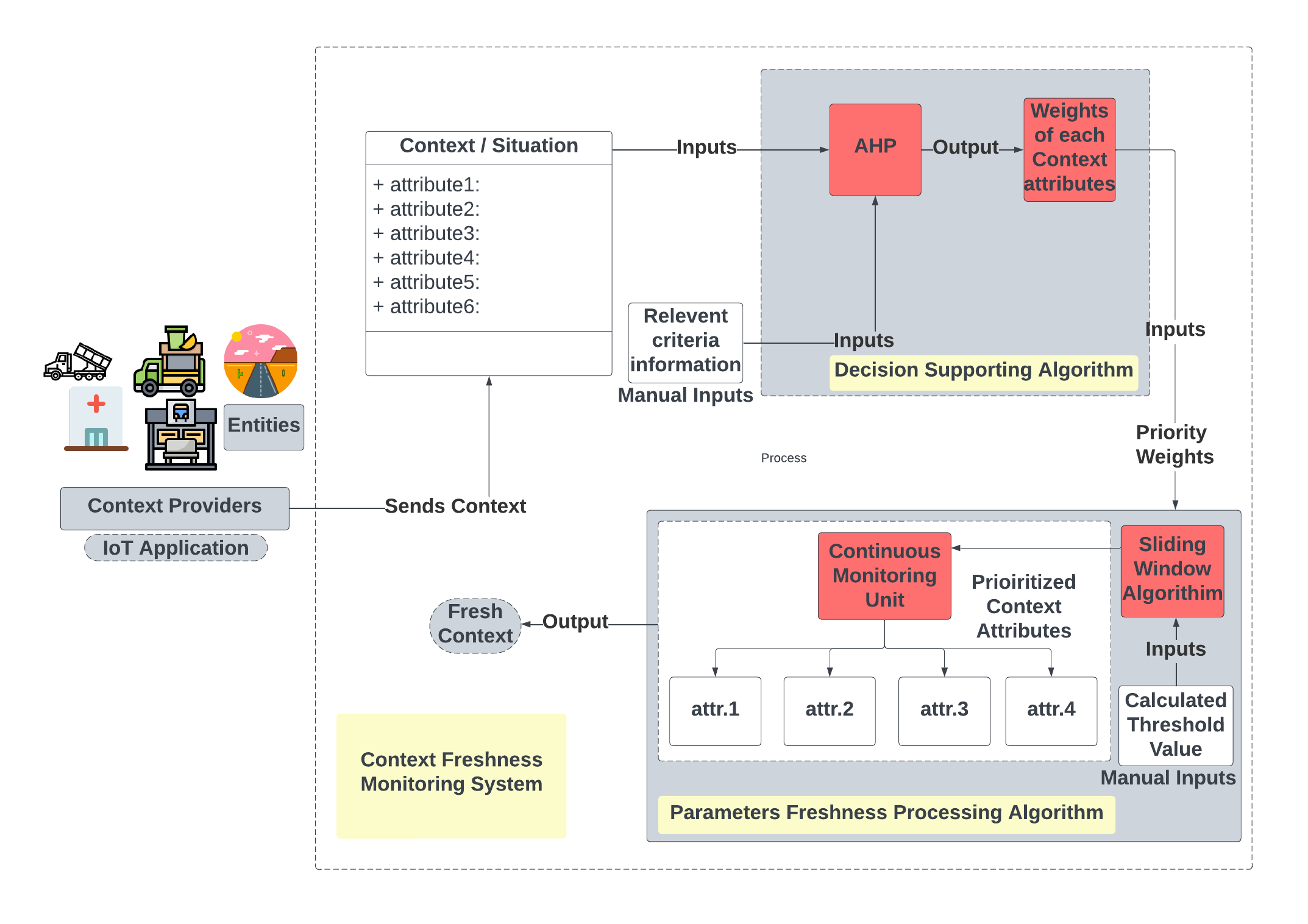}
\caption{Context Freshness Monitoring System.
} \label{fig77}
\end{figure}

Figure~\ref{fig77} demonstrates how the CFMS system works which can be incorporated into a Context Management Platform (CMP).

In the proposed system, the contextual information with all the context attributes is received from IoT applications/context providers. The system incorporates two main algorithms. Firstly, the Decision Supporting Algorithm (DSA) explained in section~\ref{sec:DesignAlg1}, extracts context attributes and forwards them to the Analytic Hierarchy Process (AHP) along with the relevant criteria information. Based on this, DSA calculates the weight of each context attribute. Once the weights are determined, the priority of the context attributes is ascertained. This prioritized information is then passed on to the Sliding Window Algorithm, which is an integral part of the Parameter Freshness Processing Algorithm (PFPA) explained in section~\ref{sec:DesignAlg2}. Within PFPA, the Sliding Window Algorithm processes the inputs received from DSA and a predefined threshold value. The output of this is then channeled to a Continuous Monitoring Unit for the cache module, which scrutinizes each parameter or context attribute against the given threshold. Subsequent caching decisions are then formulated based on observations from the continuous monitoring unit.

In managing these operations, the system, labeled as CFMS, directly monitors a key metric called ``context freshness." This metric indicates the recency or timeliness of the context information. As a result, the output from the continuous monitoring unit ensures that the context information stored in the cache remains consistently fresh.

\subsection{Decision Supporting Algorithm}~\label{sec:DesignAlg1}

The aim of this algorithm is to consider the entire context as fresh if the most essential context attributes are updated. To identify these pivotal context attributes, it's necessary to determine the weight (indicating its relevance to the context) of each attribute, achieved using the AHP method.

For the application of the Analytical Hierarchical Process (AHP)~\cite{afshari2010simple} to any given situation, the situation/context is perceived as the ``goal" or the ultimate object of consideration. The distinct context attributes of the contextual information form the criteria, which are the parameters under examination in the context. These context attributes are defined as a set CA, such that CA = {ca1, ca2, ..., can} where n represents the total number of context attributes. Each context attribute or criterion cai is then weighed and ranked through AHP.

AHP uses a pairwise comparison matrix to determine the relative importance or ``weight" of each context attribute which is then compared with every other context attribute, forming a set of n(n-1)/2 pairwise comparisons. This process results in a weight wi associated with each context attribute ci, forming a weighted set W = {w1, w2, ..., wn}. Once these weights are determined, a consistency check is performed using the consistency index (CI) and consistency ratio (CR). Provided the CR falls within an acceptable threshold, affirming consistent pairwise comparisons, the global priority weights are calculated.

A ranking is assigned to each context attribute based on its global priority weight, forming a ranked set R = {r1, r2, ..., rn}, where r1, r2, ..., rn represent the ranks of the context attributes ca1, ca2, ..., can respectively. The ranking ranges from 1 to n, with 1 being assigned to the attribute of highest priority or importance and n to the one with the least. The entire process provides an objective and quantified way to identify the key parameters for any context. It is generalizable and capable of handling 'n' number of context attributes as shown in Algorithm~\ref{algo}.

\begin{algorithm}
\caption{Decision Supporting Algorithm}

\begin{algorithmic}[1] 
\Procedure {AHP\_Decision}{Goal, Criteria} 
\State $\text{hierarchy} = \text{{``Goal": Criteria}}$ 

\State $\text{comparison\_matrix} = \text{{pairwise\_comparison(Criteria)}}$ 

\State $\text{comparison\_matrix} = \text{{build\_comparison\_matrix(Criteria, comparison\_matrix)}}$ 
\State $\text{priority\_weights} = \text{{calculate\_priority\_weights(comparison\_matrix)}}$ 
\State $CI = \text{{calculate\_consistency\_index(comparison\_matrix, priority\_weights)}}$ 
\State $ACI = \text{{calculate\_average\_consistency\_index(CI, num\_criteria)}}$ 
\State $CR = \text{{calculate\_consistency\_ratio(ACI, RI)}}$ 

\If{$CR > \text{threshold}$} 
    \State \Return $\text{``Inconsistent pairwise comparisons"}$
\EndIf

\State $\text{global\_weights} = \text{{calculate\_global\_priority\_weights(priority\_weights, hierarchy)}}$ 
\State $\text{sensitivity} = \text{{perform\_sensitivity\_analysis(comparison\_matrix, global\_weights)}}$ 
\State $\text{ranking} = \text{{rank\_criteria(global\_weights)}}$ 

\State \Return ranking
\EndProcedure
\label{algo}
\end{algorithmic}
\end{algorithm}

\begin{figure}
\centering
\includegraphics[width=12cm]{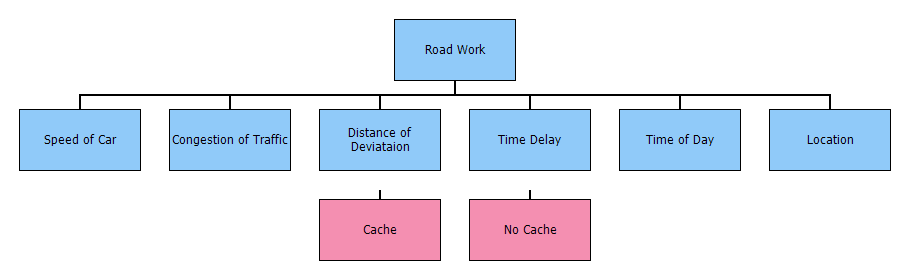}
   \caption{Hierarchy of context ``Road Work" for decision making.}
   \label{fig55}
\end{figure}

The resulting pairwise comparison matrix is then processed to calculate the weight of each context attribute. This operation is performed for all context attributes as parameters and subsequently selects the top four parameters based on their calculated weights. These chosen parameters are then cached for quick reference and used as triggers for the Sliding Window Algorithm, ensuring the context freshness of cached context. Figure~\ref{fig55} displays the hierarchy of context ``road work" with related criteria as parameters and weighing decisions of cache and no cache as alternatives.

\subsection{Parameters Freshness Processing Algorithm}~\label{sec:DesignAlg2}

The Parameters Freshness Processing Algorithm with Sliding Window Algorithm(SWA) and Continuous Monitoring Unit(CMU) operates in tandem with the outputs from the DSA. The purpose of this algorithm is to ensure that prioratized context attributes remain fresh and updated according to a specified threshold inside the cache.

\begin{algorithm}
\centering
\caption{Parameters Freshness Processing Algorithm with SWA, CMU, and Cache}

\begin{algorithmic}[1] 
\Require prioritized\_attributes from DSA \Comment{Ranked context attributes}
\Require Threshold \( T \)
\State Initialize a deque \textit{window} to hold the sliding window of data with size \textit{window\_size}
\State Initialize a set \textit{cache} to hold cached data points
\State Initialize Continuous Monitoring Unit (CMU) with prioritized\_attributes

\Procedure {SWA}{data\_point}
    \State Add data\_point to window

    \While {CMU.check\_threshold(data\_point) \textgreater{} \( T \)}
        \State old\_data\_point = window.popleft()

        \If {old\_data\_point not in window}
            \State CMU.update\_context\_attribute(old\_data\_point)
            \State Remove old\_data\_point from cache
        \EndIf
    \EndWhile
\EndProcedure

\For {each data\_point in prioritized\_attributes}
    \If {data\_point not in cache}
        \State Add data\_point to cache
    \EndIf
    \State \Call{SWA}{data\_point}
    \If {CMU.check\_threshold(data\_point) \textgreater{} \( T \)}
        \State Update cache with the fresh value of data\_point
    \EndIf
\EndFor
\end{algorithmic}
\label{algo2}
\end{algorithm}

\subsubsection{Inputs and Initializations}

The algorithm requires two main inputs:
\begin{itemize}
    \item \textbf{prioritized\_attributes}: This is a list of context attributes ranked based on their weights, as determined by the DSA.
    \item \textbf{Threshold \( T \)}: A predefined threshold value which is used to determine when a context attribute value should be updated or removed from the sliding window.
\end{itemize}

Upon initiation:
\begin{enumerate}
    \item A deque, named \textbf{window}, is initialized to act as the sliding window, which holds context attributes. The size of this window is governed by the variable \textbf{window\_size}.
    \item The CMU is initialized with the prioritized context attributes obtained from DSA.
\end{enumerate}

\subsubsection{Sliding Window Algorithm (SWA) Procedure}

This procedure processes individual data points by:
\begin{enumerate}
    \item Adding the context attribute value to the window.
    \item Checking if the current value's freshness or value surpasses the threshold \( T \) as per the CMU's evaluation. If it does:
    \begin{enumerate}
        \item The oldest value is removed from the window.
        \item If this removed value is no longer present in the current window, the CMU updates its records to reflect the removal of this context attribute.
    \end{enumerate}
\end{enumerate}

\subsubsection{Main Loop}

For each value in the prioritized list of context attributes, the algorithm calls the SWA procedure. This ensures all attributes are checked and the sliding window is adjusted accordingly to maintain context freshness.

This algorithm thereby provides a systematic approach to handle and update context attributes based on their weights and freshness, ensuring that the most significant attributes are always up-to-date and maintained.

\section{Experimental Evaluation}~\label{sec:evaluation}

This section introduces an experimental evaluation designed to ascertain the efficiency of the caching mechanism in maintaining the context freshness by monitoring the context attributes identified through the Analytical Hierarchical Process (AHP). The effectiveness of the CFMS is measured using \textbf{cache hit} and \textbf{cache miss} ratios as key performance indicators, illuminating the efficiency of the caching system. Furthermore, a separate metric accounting for expired items in the cache - \textbf{cache expired} ratio - is introduced to specifically measure the context freshness aspect. The experimental setup is being tweaked in three primary ways:

\begin{enumerate}
    \item Varying the threshold for the Sliding Window Algorithm: This manipulation allows the determination of the point at which data crosses the line from being fresh to being ``stale". The identification of this threshold is critical for context freshness preservation.
    
    \item Varying the volume of entries/contextual information entering the system from various IoT applications: This variation enables an evaluation of how effectively the system can handle different levels of demand, and whether or not this impacts context freshness and scalability.
    
    \item Altering the cache capacity: By adjusting the scaling up or down the context memory as in~\cite{10.1145/3558053.3558057}, the influence of storage capacity on the system's performance and the context freshness as key metric of the cached context was tested.
\end{enumerate}

\begin{figure}
\centering
\includegraphics[width=8cm]{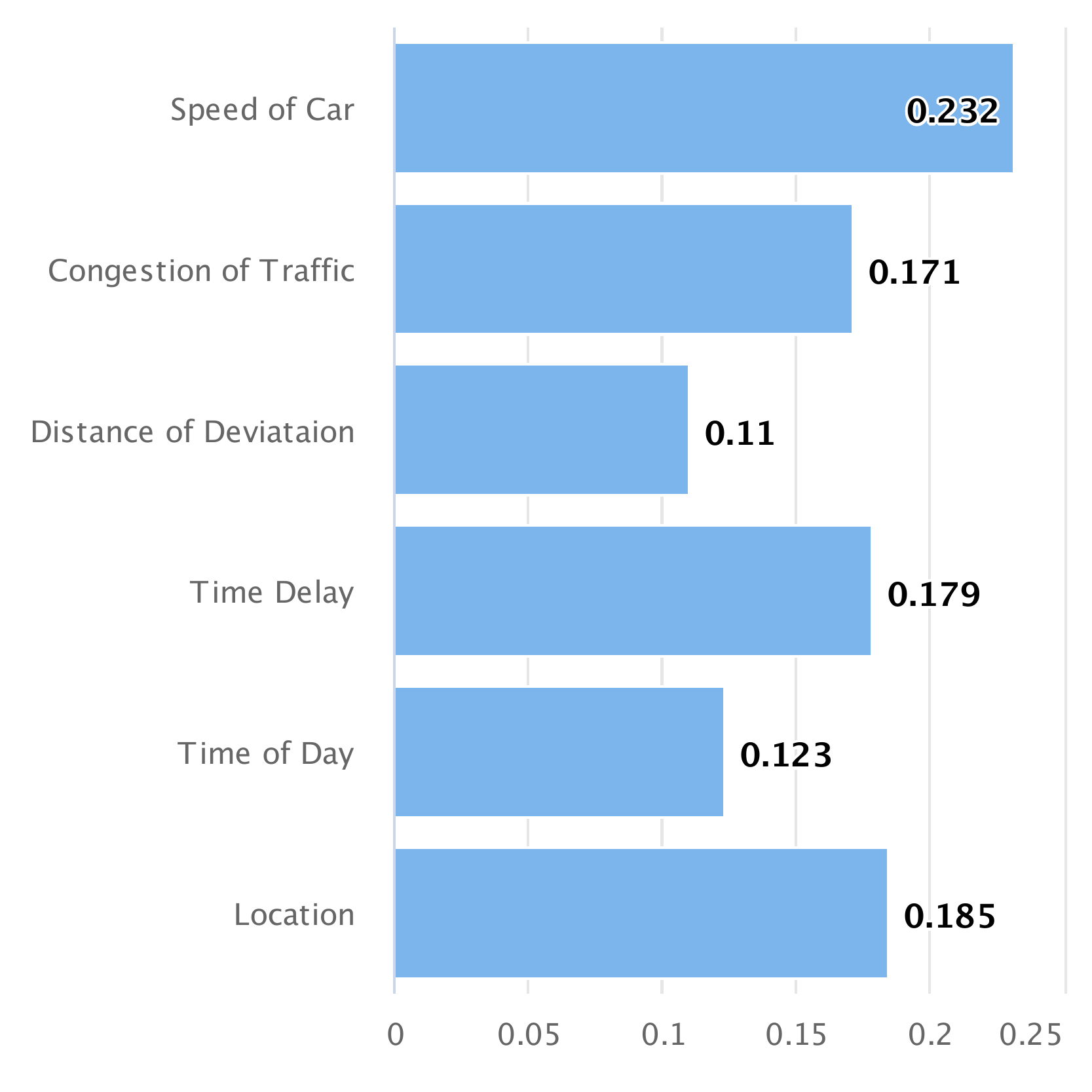}
\caption{Weights of context attributes of context ``Road Work" after DSA.}
\label{fig3}
\end{figure}

Expanding upon the Analytical Hierarchical Process (AHP) methodology described in Section~\ref{sec:DesignAlg1}, the computed weights are illustrated in Figure~\ref{fig3}. These weights mark the priority of the different criteria related to the context ``road work", indicating which context attributes are most important and thus guiding the caching process within the PFPA. This process, coupled with continuous monitoring, ensures the maintenance of context freshness.

The outputs generated from the DSA are subsequently used as inputs to monitor the context freshness of context attributes(parameters) via the Algorithm~\ref{algo2}. This algorithm primarily focuses on maintaining a sliding window of parameters and tracking the performance of caching in terms of cache hit-and-miss ratios. 

\subsection{Varying the threshold for the Sliding Window Algorithm}

In this subsection, the performance of the caching system is evaluated in terms of cache hit and cache miss ratios, with the key variable being the threshold value set for the ``sliding window algorithm", which means after the threshold is reached, the IoT data corresponding to the context attribute will be considered as stale and evicted from cache. The threshold is systematically varied from 10 minutes to 25 minutes, in increments of 5 minutes, as indicated in Table~\ref{tab1}.

\begin{table}[H]
    \centering
    \setlength{\tabcolsep}{8pt} 
    \renewcommand{\arraystretch}{1.5} 
    \begin{tabular}{|l|l|l|l|l|}
    \hline
        \multirow{2}{*}{Threshold} & \multicolumn{4}{c|}{Value} \\ \cline{2-5}
        & 10 & 15 & 20 & 25 \\ \hline
        Cache hit & 174 & 179 & 184 & 186 \\ \hline
        Cache miss & 26 & 21 & 16 & 14 \\ \hline
        Ratio & 6.7 & 8.5 & 11.5 & 13.3 \\ \hline
    \end{tabular}
    \vspace{10pt} 
    \caption{Cache hit and cache miss ratios at different thresholds.}
    \label{tab1}
\end{table}

\begin{figure}
\centering
\includegraphics[width=12cm]{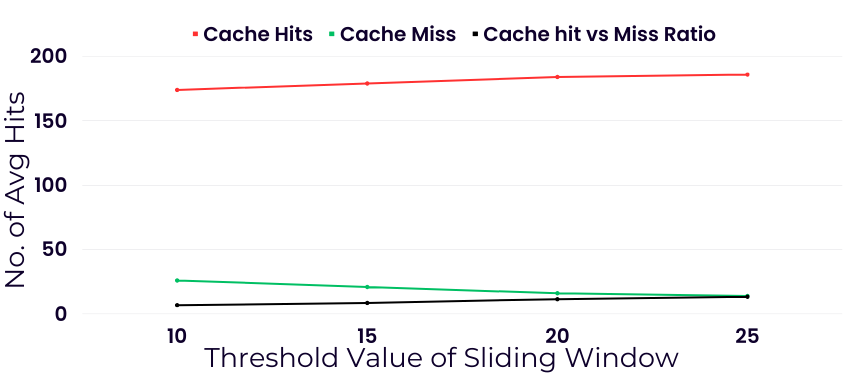}
\caption{Comparison of Cache hit with a threshold value of Sliding Window.
} \label{fig4}
\end{figure}

The results of this variation, visualized in Figure~\ref{fig4}, suggest a trend of increasing cache hits as the threshold value rises. After analysis, a 20-minute threshold has been selected for the experiments conducted in the subsequent sections. It's important to note that post a threshold of 22 minutes, no significant impact or changes were observed in the system's performance. This threshold selection ensures an optimal balance between cached context freshness and computational efficiency.

\subsection{Changing the size of the incoming entries(contextual data)}

In this subsection, the size of the incoming entries load was systematically varied, testing with 150, 250, 350, and 500 for each threshold from 10 to 25 minutes in increments of 5 minutes as shown in Table~\ref{tab2}. The findings reveal a consistent pattern across all test cases. With an increasing number of entries, both cache hit and cache miss counts increase, but the cache hit ratio remains relatively consistent indicating that the ``hybrid approach" also supports scalability.

\begin{table}[H]
    \centering
    \setlength{\tabcolsep}{8pt} 
    \renewcommand{\arraystretch}{1.5} 
    \begin{tabular}{|l|l|l|l|l|}
    \hline
        No. of queries & Threshold & Cache Hit & Cache miss & Cache Hit Ratio \\ \hline
        \multirow{4}{*}{150} & 10 & 528 & 72 & 7.33 \\ \cline{2-5}
        & 15 & 542 & 58 & 9.34 \\ \cline{2-5}
        & 20 & 555 & 45 & 12.33 \\ \cline{2-5}
        & 25 & 561 & 39 & 14.38 \\ \hline
        \multirow{4}{*}{250} & 10 & 880 & 120 & 7.33 \\ \cline{2-5}
        & 15 & 904 & 96 & 9.41 \\ \cline{2-5}
        & 20 & 925 & 75 & 12.33 \\ \cline{2-5}
        & 25 & 934 & 66 & 14.15 \\ \hline
        \multirow{4}{*}{350} & 10 & 1232 & 168 & 7.33 \\ \cline{2-5}
        & 15 & 1266 & 134 & 9.44 \\ \cline{2-5}
        & 20 & 1296 & 104 & 12.46 \\ \cline{2-5}
        & 25 & 1309 & 91 & 14.38 \\ \hline
        \multirow{4}{*}{500} & 10 & 1761 & 239 & 7.36 \\ \cline{2-5}
        & 15 & 1809 & 191 & 9.47 \\ \cline{2-5}
        & 20 & 1851 & 149 & 12.42 \\ \cline{2-5}
        & 25 & 1870 & 130 & 14.38 \\ \hline
    \end{tabular}
    \vspace{10pt} 
    \caption{Cache hit and cache miss ratios at different number of entries.}
    \label{tab2}
\end{table}

\begin{figure}
\centering
\includegraphics[width=10cm]{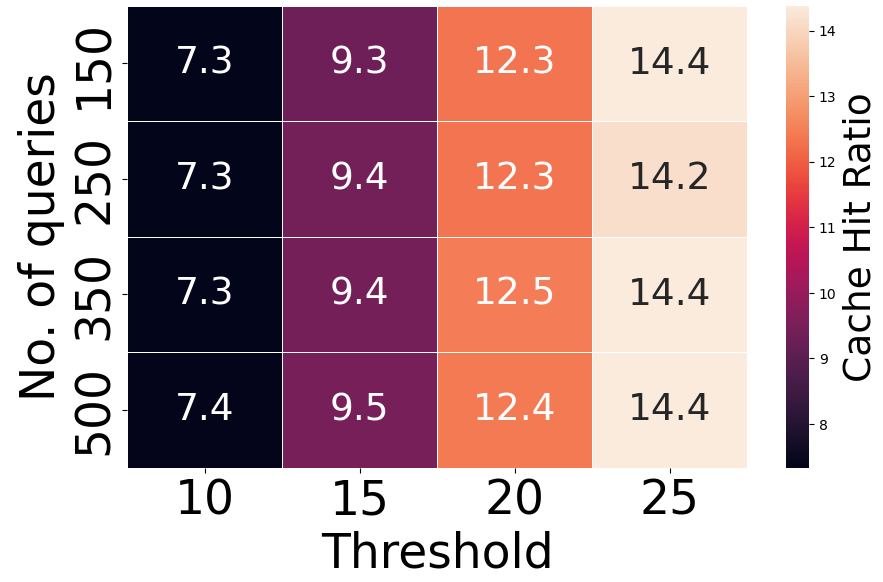}
\caption{Comparison of Cache hit ratio with Number of Entries and Thresholds.
} \label{fig5}
\end{figure}

From the 'Cache Hit Ratio' heatmap (Figure~\ref{fig5}), we can observe a pattern of increasing cache hit ratio with an increasing threshold for all entry sets. For a threshold of 10, the cache hit ratio remains relatively steady around 7.33 to 7.36 across all entries. As the threshold increases to 15, there is a notable improvement in the ratio, reaching up to 9.47 for 500 entries. When the threshold is increased further to 20 minutes, the ratio experiences an additional boost to a range of approximately 12.33 to 12.46. Interestingly, upon reaching a 25-minute threshold, the ratio increases to around 14.38 for all query sets, except for 250 queries where it marginally drops to 14.15. This discrepancy could be attributed to various factors including caching policies, size of the cache, or variability in the access patterns. 

These findings, illustrated in the heatmap, affirm the choice of a 20-minute threshold as a suitable point. While the cache hit ratio generally improves with an increase in threshold, the gains beyond the 20-minute mark are relatively minor. This confirms the trade-off between context freshness and computational efficiency, and indicates the diminishing returns of increasing the threshold beyond 20 minutes. Therefore, a 20-minute threshold appears to be the optimal point for maintaining an efficient cache system, given the current configuration and workload.

\subsection{Modifying the cache capacity}

\begin{figure}
\centering
\includegraphics[width=12cm]{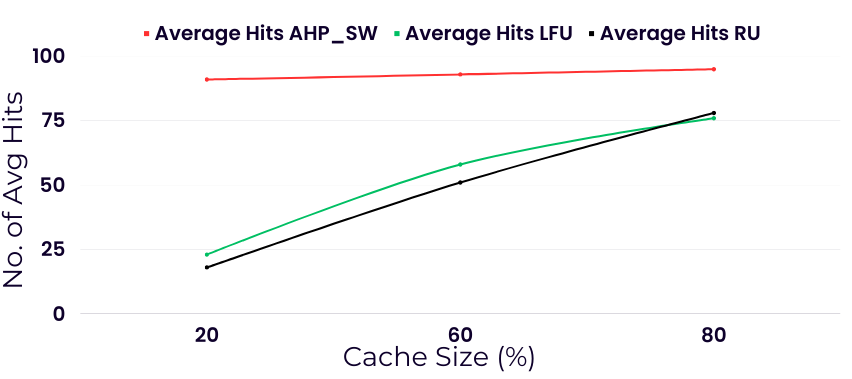}
\caption{ Comparison of the average number of cache hits for
different cache storage limit.
} \label{fig6}
\end{figure}

In this subsection, the cache capacity is adjusted to varying capacity - 20\%, 60\%, and 80\%. This test keeps the number of incoming entries constant at 500 and sets the threshold at 20 minutes. The experiment aims to compare the efficiency of using DSA \& PFPA in caching with other caching algorithms, namely LFU (Least Frequently Used) and RU (Recently Used). The results of this comparison are displayed in Figure~\ref{fig6}.

A careful analysis of the results reveals that as the cache size increases from 20\% to 80\%, the use of DSA \& PFPA experiences a slight increase in average cache hits, from 91 to 95. Comparatively, the LFU algorithm exhibits a more substantial increase in average cache hits, growing from 23 to 76 with the increase in cache size. Similarly, the RU algorithm demonstrates a significant rise in average cache hits, from 18 to 78, as the cache size increases.

These results suggest that while increasing cache capacity does enhance average cache hits for all algorithms, the use of DSA \& PFPA appears less sensitive to changes in cache capacity. This indicates more efficient utilization of cache space by using DSA \& PFPA in caching which takes into account for monitoring context attributes and maintaining the context freshness, thereby reinforcing its suitability and advantage in real-time IoT applications, where memory resources may be limited.

\subsection{Evaluating the Cache Expired Ratio}

\begin{figure}
\centering
\includegraphics[width=12cm]{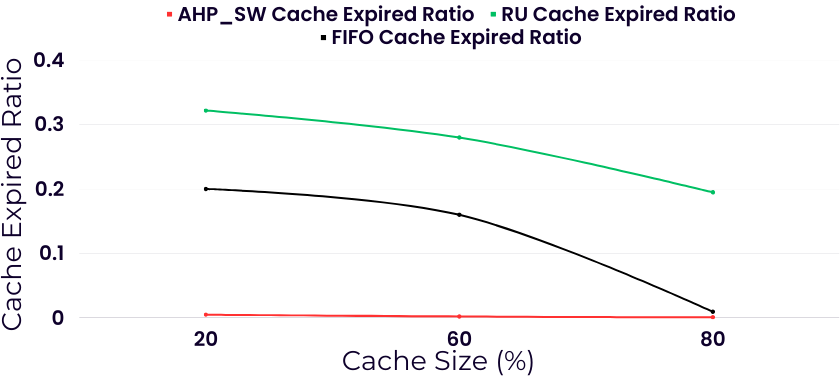}
\caption{ Comparison of the cache expired ratio for
different cache storage limit.
} \label{fig7}
\end{figure}

Figure~\ref{fig7} provides a comparative view of the cache expired ratio - a measure of context freshness metric - with two different caching algorithms: DSA \& PFPA used in caching, Recently Used (RU), and First In, First Out (FIFO). An essential observation from the figure is the distinct capability of the monitoring ability of DSA \& PFPA to perform efficiently even when the cache size is as low as 20. This significant feature underscores its potential applicability in scenarios like network edge or fog computing, where memory constraints are prevalent. As more systems aim to achieve data/process localization and real-time operations, the DSA \& PFPA's ``context freshness" monitoring proficiency at low cache sizes becomes a vital contribution of this work.

As the cache size increases from 20 to 80, the cache expired ratio calculated using DSA \& PFPA remains consistently low, highlighting its superior ability to maintain context freshness. Even with increasing cache size, this monitoring algorithm ensures storage of only the most recent and relevant context, indicating effective cache management.

Conversely, the RU algorithm, starting with a high cache expired ratio of 0.322 at a cache size of 20, shows a decrease to 0.195 as the cache size expands to 80. While this indicates some improvement in context freshness with a growing cache size, it is still less efficient than DSA \& PFPA . FIFO, which starts with a cache expired ratio of 0.2 at a cache size of 20, observes a significant drop to 0.0095 at a cache size of 80. This sharp decrease, however, may not necessarily signify high context freshness metric, especially given its initially high ratio. DSA \& PFPA establishes its robustness and efficiency by continuously monitoring parameters and maintaining the context freshness as a preferred mechanism for caching algorithm for real-time IoT applications, especially in environments with memory constraints.

\section{Discussion}~\label{sec:discussion}

The evaluation has presented significant findings. Using DSA and PFPA to monitor parameters and maintaining the metric of ``context freshness" has improved context caching. One important observation was made during the threshold experiment. The cache hit rate increased as the threshold value went up. However, after reaching the 20-minute point, there was only a small improvement. This highlights the importance of finding a balance between maintaining context freshness and optimizing computational efficiency. It's essential to choose a threshold that not only increases hit rates but also preserves context freshness, especially for real-time IoT applications. Based on the evaluation, a 20-minute threshold appears to provide this balance.

Additionally, altering the amount or number of contextual information entries into the CFMS demonstrated the strong scalability of the proposed hybrid approach. As the number of entries increases, the proportion of cache hits stayed steady because increasing the entries didn't affect the CFMS. The system continued to monitor the parameters linked to the new entries. This consistency, even with changing entries, supports the system's reliability, making it appropriate for various IoT applications. Yet, the most notable observation came from the changes in cache capacity. While all other algorithms (RU, LFU) performed better with larger cache sizes, but the performance of caching context using DSA \& PFPA maintaining the context freshness was less dependent on increasing the cache size. This adaptability to size variations can be very useful in time-sensitive IoT applications, particularly in edge computing environments where memory may be limited.

Finally, the cache expired ratio, which monitors the metric ``context freshness", showed that DSA \& PFPA performed notably better. Even with smaller cache sizes, the CFMS effectively kept the cache expired ratio low. This capability is important for real-time IoT applications, especially in environments with limited memory restrictions. This observation highlights the potential of DSA \& PFPA as a promising approach, suggesting more research and testing in different IoT environments.

\section{Conclusion \& Future Work}

The study indicates that by monitoring parameters by maintaining metric ``context freshness", which can keep the context fresh can significantly improve the performance of caching in real-time context-aware IoT applications. The system, which uses DSA \& PFPA, monitors the prioritized parameters to ensure the context remains fresh, leading to a stable cache hit rate. This hybrid approach provides a dynamic response to environmental changes, ensuring that both the freshness of the cached context and the system's overall efficiency are enhanced.

Looking forward, there are several exciting possibilities for further enhancing of CFMS. One such improvement could be the integration of fuzzy AHP, which would provide a more nuanced understanding of the relative importance of each criterion/parameter. The incorporation of decision trees could further refine this approach, enabling more complex and layered decision-making. Plans are also underway to incorporate this enhanced system into a broader context management platform like CoaaS.

\section{Acknowledgement}
Support for this publication from the Australian Research Council (ARC) Discovery Project Grant DP200102299 is thankfully acknowledged.


\begin{thebibliography}{8}

\bibitem{s23031711}
Manchanda, A., Lee, K., Poznanski, G.D., Hassani, A.: Automated Adjustment of PPE Masks Using IoT Sensor Fusion. In: Sensors, vol. 23, no. 3, article no. 1711. (2023)


\bibitem{al2015internet}
Al-Fuqaha, A., Guizani, M., Mohammadi, M., Aledhari, M., Ayyash, M.: Internet of Things (IoT): A vision, architectural elements, and future directions. In: Future Generation Computer Systems, vol. 39, pp. 164--176. Elsevier (2015)

\bibitem{ryan2000from}
Ryan, N., Pascoe, J.: From context-aware to context-driven: a paradigm shift. In: Proceedings of the workshop on situated interaction in ubiquitous computing, September 2000.



\bibitem{forkan2022mobile}
Forkan, A. R. M., Kang, Y.-B., Marti, F., Joachim, S., Banerjee, A., Milovac, J. K., Jayaraman, P. P., McCarthy, C., Ghaderi, H., Georgakopoulos, D.: Mobile IoT-RoadBot: An AI-Powered Mobile IoT Solution for Real-Time Roadside Asset Management. In: Proceedings of the 28th Annual International Conference on Mobile Computing And Networking, pp. 883--885. Association for Computing Machinery, New York, NY, USA (2022)




\bibitem{dey2001understanding}
Dey, A. K.: Understanding and using context. \textit{Personal and Ubiquitous Computing}, 5 (2001), pp. 4--7.




\bibitem{hassani2018context}
Hassani, A., Medvedev, A., Haghighi, P. D., Ling, S., Indrawan-Santiago, M., Zaslavsky, A., Jayaraman, P. P.: Context-as-a-Service Platform: Exchange and Share Context in an IoT Ecosystem. In: Proceedings of the 2018 IEEE International Conference on Pervasive Computing and Communications Workshops (PerCom Workshops), pp. 385--390. doi:10.1109/PERCOMW.2018.8480240






\bibitem{perera2014context}
Perera, C., Zaslavsky, A., Christen, P., Georgakopoulos, D.: Context aware computing for The Internet of Things: A survey. In: IEEE Communications Surveys \& Tutorials, vol. 16, no. 1, pp. 414--454. IEEE (2014)

\bibitem{KRASNIQI2016269}
Krasniqi, X., Hajrizi, E.: Use of IoT Technology to Drive the Automotive Industry from Connected to Full Autonomous Vehicles. In: IFAC-PapersOnLine, vol. 49, no. 29, pp. 269-274. Elsevier (2016)

\bibitem{Iyer2021AI}
Iyer, L.S.: AI enabled applications towards intelligent transportation. In: Transportation Engineering, vol. 5, 2021, 100083, ISSN 2666-691X. Elsevier (2021)




\bibitem{Al-Ward2022Caching}
Al-Ward, H., Tan, C.K., Lim, W.H.: Caching transient data in Information-Centric Internet-of-Things (IC-IoT) networks: A survey. In: Journal of Network and Computer Applications, vol. 206, 2022, 103491, ISSN 1084-8045. Elsevier (2022)



\bibitem{khargharia2022probabilistic}
Khargharia, H.: Probabilistic analysis of context caching in the Internet of Things applications. In: IEEE Xplore (2022). Available at: \url{https://ieeexplore-ieee-org.ezproxy-b.deakin.edu.au/document/9860212}




\bibitem{weerasinghe2023traditional}
Weerasinghe, S., Zaslavsky, A., Loke, S. W., Hassani, A., Abken, A., Medvedev, A.: From Traditional Adaptive Data Caching to Adaptive Context Caching: A Survey. arXiv preprint arXiv:2211.11259 (2023)

\bibitem{10.1145/3558053.3558057}
S.~Weerasinghe, A.~Zaslavsky, S. W.~Loke, A.~Medvedev, and A.~Abken,
\textit{Estimating the Dynamic Lifetime of Transient Context in near Real-Time for Cost-Efficient Adaptive Caching},
SIGAPP Appl. Comput. Rev., 
vol.~22, no.~2, pp.~44–58, Aug.~2022.





\bibitem{bettini2010context}
Bettini, C., Brdiczka, O., Henricksen, K., Indulska, J., Nicklas, D., Ranganathan, A., Riboni, D.: A survey of context modelling and reasoning techniques. arXiv preprint arXiv:1611.01800 (2010)



\bibitem{stankovic2014research}
Stankovic, J. A.: Research Directions for the Internet of Things. arXiv preprint arXiv:1401.0003 (2014)

\bibitem{gu2005middleware}
Gu, T., Pung, H. K., Zhang, D. Q.: A middleware for building context-aware mobile services. arXiv preprint arXiv:0512.2656 (2005)


\bibitem{rizwan2016realtime}
Rizwan, P., Suresh, K., Babu, M. R.: Real-time smart traffic management system for smart cities by using Internet of Things and big data. In: Proceedings of the 2016 International Conference on Emerging Technological Trends (ICETT), pp. 1--7. doi:10.1109/ICETT.2016.7873660



\bibitem{shekade2020vehicle}
Shekade, A., Mahale, R., Shetage, R., Singh, A., Gadakh, P.: Vehicle Classification in Traffic Surveillance System using YOLOv3 Model. In: Proceedings of the 2020 International Conference on Electronics and Sustainable Communication Systems (ICESC), pp. 1015--1019. doi:10.1109/ICESC48915.2020.9155702



\bibitem{fernandez2019realtime}
Fernández-Sanjurjo, M., Bosquet, B., Mucientes, M., Brea, V. M.: Real-Time Visual Detection and Tracking System for Traffic Monitoring. \textit{Engineering Applications of Artificial Intelligence}, doi:10.1016/j.engappai.2019.07.005 (2019)





\bibitem{Muller2017}
S.~Müller, O.~Atan, M.~van der Schaar, and A.~Klein,
\textit{Context-Aware Proactive Content Caching With Service Differentiation in Wireless Networks},
IEEE Transactions on Wireless Communications, 
vol.~16, no.~2, pp.~1024-1036, Feb.~2017.

\bibitem{XU201924}
C.~Xu, and X.~Wang,
\textit{Transient content caching and updating with modified harmony search for Internet of Things},
Digital Communications and Networks, 
vol.~5, no.~1, pp.~24-33, 2019.


\bibitem{app8101959}
T.~Fang, H.~Tian, X.~Zhang, X.~Chen, X.~Shao, and Y.~Zhang,
\textit{Context-Aware Caching Distribution and UAV Deployment: A Game-Theoretic Approach},
Applied Sciences, 
vol.~8, no.~10, Article no.~1959, 2018.





\bibitem{ref_article1}
Mcheick, H., Khreis, M., Al-Kalla, M., Sweidan, H.: CARE Framework: Context-Aware Reliable Engine Health Focus on Traffic Monitoring System. In: 2014 UKSim-AMSS 16th International Conference on Computer Modelling and Simulation, pp. 411-416. March 2014. doi:10.1109/UKSim.2014.76



\bibitem{ref_article2}
Goel, D., Pahal, N., Jain, P., Chaudhury, S.: An ontology-driven context aware framework for smart traffic monitoring. In: 2017 IEEE Region 10 Symposium (TENSYMP), pp. 1-5. 2017. doi:10.1109/TENCONSpring.2017.8070059


\bibitem{ref_article3}
Serdaroglu, K. C., Baydere, S.: On the Data Freshness for IoT Traffic Modelling in Real-Time Emergency Observation Systems. In: 2018 23rd Conference of Open Innovations Association (FRUCT), pp. 335-340. 2018. doi:10.23919/FRUCT.2018.8588029

\bibitem{ref_article4}
Faro, A., Giordano, D., Spampinato, C.: Evaluation of the Traffic Parameters in a Metropolitan Area by Fusing Visual Perceptions and CNN Processing of Webcam Images. \textit{IEEE Transactions on Neural Networks} \textbf{19}(6), 1108--1129 (2008). doi:10.1109/TNN.2008.2000392

\bibitem{ref_article5}
A. Pande and M. Abdel-Aty, ``Assessment of freeway traffic parameters leading to lane-change related collisions," \textit{Accident Analysis \& Prevention}, vol. 38, no. 5, pp. 936-948, 2006. doi:10.1016/j.aap.2006.03.004


\bibitem{ref_article6}
Beymer, D., McLauchlan, P., Coifman, B., Malik, J.: A real-time computer vision system for measuring traffic parameters. In: Proceedings of IEEE Computer Society Conference on Computer Vision and Pattern Recognition, pp. 495-501. 1997. doi:10.1109/CVPR.1997.609371



\bibitem{ref_article7}
Thakur, T. T., Naik, A., Vatari, S., Gogate, M.: Real-time traffic management using Internet of Things. In: 2016 International Conference on Communication and Signal Processing (ICCSP), pp. 1950-1953. 2016. doi:10.1109/ICCSP.2016.7754512


\bibitem{padovitz2004theory}
Padovitz, A., Loke, S.W., Zaslavsky, A.: Towards a theory of context spaces. In: Proceedings of the Second IEEE Annual Conference on Pervasive Computing and Communications Workshops, pp. 38--42 (2004). doi:10.1109/PERCOMW.2004.1276902

\bibitem{boytsov2011sensory}
Boytsov, A., Zaslavsky, A.: From Sensory Data to Situation Awareness: Enhanced Context Spaces Theory Approach. In: Proceedings of the 2011 IEEE Ninth International Conference on Dependable, Autonomic and Secure Computing, pp. 207--214 (2011). doi:10.1109/DASC.2011.55


\bibitem{hassani2019context}
Hassani, A., Medvedev, A., Delir Haghighi, P., Ling, S., Zaslavsky, A., Prakash Jayaraman, P.: Context Definition and Query Language: Conceptual Specification, Implementation, and Evaluation. Sensors (Basel), 19(6), pp. 1478 (2019). PMID: 30917602; PMCID: PMC6470624.


\bibitem{afshari2010simple}
Afshari, A., Mojahed, M., Yusuff, R.: Simple Additive Weighting Approach to Personnel Selection Problem. International Journal of Innovation, Management and Technology, 1, pp. 511-515 (2010)doi:10.7763/IJIMT.2010.V1.89.








\end{thebibliography}
\end{document}